\documentclass{JAC2000}
\usepackage{graphicx}

\begin{document}
\title{\flushright{WEDT002}\\[15pt] \centering THE KLOE/DA$\Phi$NE STATUS LOGGING, ANALYSIS AND DATABASE SYSTEM}

\author{G.~Mazzitelli, F.~Murtas, P.~Valente\\
Laboratori Nazionali di Frascati - INFN, Frascati (Roma), I-00044, Italy\\
}
\maketitle

\begin{abstract} 
The KLOE experiment~\cite{kloe} at 
the Frascati $\phi$-factory DA$\Phi$NE~\cite{dafne}, 
designed to measure $\Re(\varepsilon^\prime/\varepsilon)$,
began preliminary data taking in the Fall of 1999. 
A large database structure, which logs information coming from 
the DA$\Phi$NE control system and the KLOE slow control and data 
acquisition systems, has been developed. 
Data from detector monitoring, online event processing and machine 
operating conditions are easily accessible for online and offline analysis 
by means of Web tools and histogramming tools.
The system allows powerful real-time data correlations which 
are necessary for the ongoing program of luminosity and background
improvements. Data flow and handling processes are presented.
\end{abstract}

\section{DA$\Phi$NE control system}
Data in the DA$\Phi$NE collider are stored by the control 
system~\cite{dafnecontrol} in the local memories
of the 45 front-end VME-CPU's distributed all over the accelerator area. 
The front-end tasks get commands from the high-level user environment and 
continually update their own database with information from the
devices. 
Data are available through direct memory access to the CPU's memory. 
This front-end database is constituted by different data types tailored to 
specific machine elements (non-homogenous database); 
this means that in order to use this data or to correlate parameters of 
different devices, specific routines must be implemented.

Two system tasks have been developed in order to collect all the parameters 
(Dumper), to synchronize and align them, and then to store them to disk 
(Storer)~\cite{datadafne}. 
The Dumper continuously fetches data from the front-end memories and 
writes the different data-types from each machine element aligning them 
in a homogenous database. 
This memory resident database is accessible from high-level
tasks:
\begin{Itemize}
\item for monitoring purposes, such as the watchdog process
checking for faulty magnets, bad vacuum or CPU failures; 
\item for the user interface display (from any operator console);
\item for online correlation and analysis (as described in Sect.~\ref{common}).
\end{Itemize} 
The Storer then synchronizes the memory database and writes it on the
mass storage at given time intervals. 

\begin{figure*}[htb]
\centering
\includegraphics*[width=0.9\textwidth]{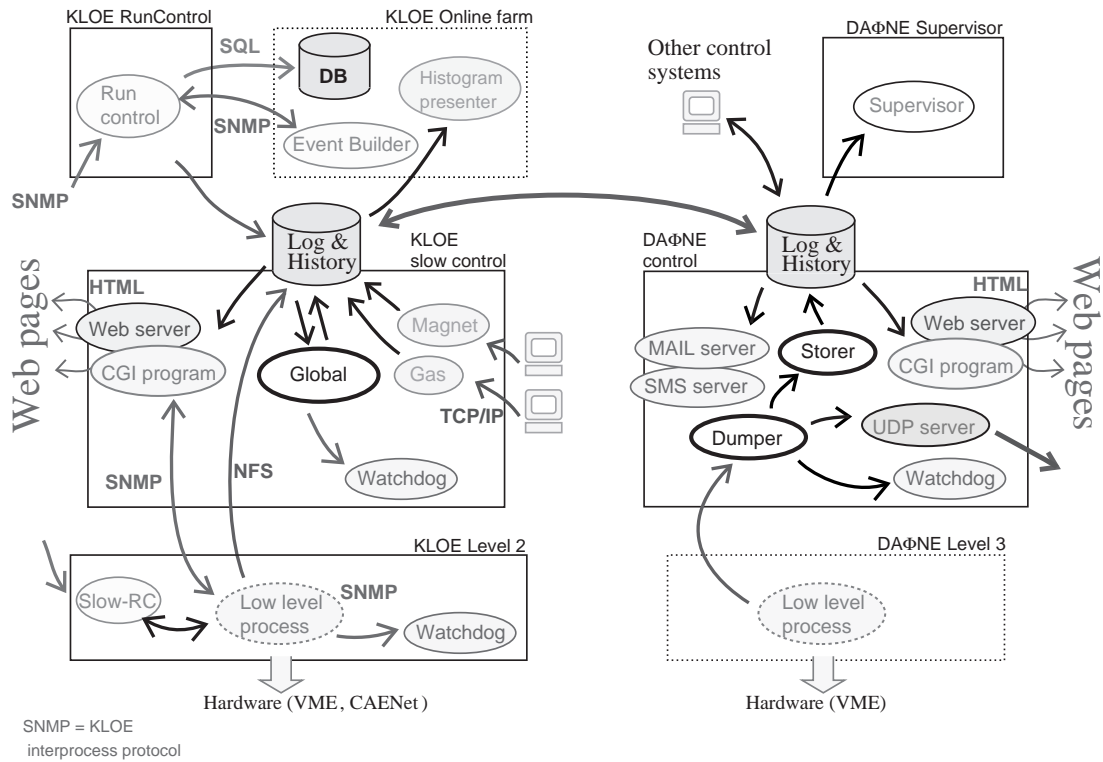}
\caption{Layout of the KLOE/DA$\Phi$NE controls and of the joint 
logging and database system.}
\label{scheme}
\end{figure*}

In addition, the Dumper continously reads the status of the satellite systems not
belonging to the control system environment (such as the spectrum analyzer), 
merging this data in the homogenous database. 
In the general scheme of DA$\Phi$NE controls,  
sketched in the right part of Fig.~\ref{scheme},
the main processes and data flow are reported. In addition
to the control and monitoring functions, and the communication with
the KLOE slow control, an additional process serves
DA$\Phi$NE data to the controls
of the synchrotron radiation facility through the UDP protocol
(UDP server). The handling of 
data relative to the DEAR experiment (running at the second interaction region)
is not shown in the figure, since it is monitored as any other part of the 
machine and the data can be presented through the DA$\Phi$NE interface.

\section{KLOE slow control}
The KLOE slow control system~\cite{kloeslow} is intended not only for the 
control and monitoring of the low and high voltages used for the detector 
(more than 70 DAQ crates, 24 HV crates), but also for the monitoring 
of the trigger and background rates. The hardware is entirely based on 
the VME standard. Several VME serial interface boards control the HV and LV 
settings and the power supplies through low-level processes running on 
a standard KLOE level-2 VME CPU.

Different low-level processes (one for each part of the detector) 
implement the low-level VME functions. All the data coming from the 
monitoring are stored to memory and are handled by high-level processes 
running on a remote machine, which also runs the processes providing the 
user interface, generating and handling alarm conditions and communicating
with satellite systems. 
The drift chamber gas control and the superconducting magnet system 
are also monitored by dedicated high level processes, communicating on 
TCP/IP sockets with the remote controls.
Finally, the high-level also communicates with the KLOE run control,
both for the logging of the relevant detector parameters, and for the setting 
of HV and LV at run start.

The user interface is \textit{entirely} realized using HTML language, 
so that the monitoring and control functions are implemented by CGI 
(Common Gateway Interface) programs running
on a dedicated Web server, which also displays the anomalous conditions
and handles the alarms. The alarm conditions are generated by two 
different watchdog processes: one running on the low-level
VME CPU and accessing the shared memories of the control processes
of the subdetectors (via SNMP, the KLOE inter-process communication
protocol~\cite{daq}); the other checking the high-level programs on the 
main machine, and sending the main alarm conditions via the GSM
short message service and e-mail.

All the slow control processes write the monitor parameters to a general 
'run condition' mass  storage. The same area is used by DAQ monitoring
processes, running on the KLOE online farm, which perform a fast event 
reconstruction of acquired data~\cite{daq}. 
These monitoring processes produce the relevant quantities from 
online analysis, such as luminosity, beam position, and background level. 

The general scheme of the KLOE controls, with the interactions of the
different low and high-level processes, occupies the left part of 
Fig.~\ref{scheme}.

\section{KLOE/DA$\Phi$NE communication and data integration}
\label{common}
The DA$\Phi$NE and KLOE mass storages, shared between the two control 
systems, are used by many monitoring processes to log a number
of parameters, in general at very different time intervals:
3 seconds for the scalers counting the hits in the endcap and 
quadrupole calorimeters, 15 seconds 
for 'fast' variables such as current and roundness of the beams, or 
the status of the low and high voltages,
1 minute for the 'slow' variables such as the KLOE magnetic field, 
5 minutes for the machine vacuum and orbit;
or even at non-constant time intervals for the physics quantities, 
such as luminosity and beam position and momentum measurements, 
needing some data to be acquired and analysed by the DAQ monitoring 
processes.
 
Thus two collaborating tasks, continously running on the two main  
machines, merge the information coming from the various parts of 
the two systems and synchronize them:
\begin{Itemize} 
\item the Dumper/Storer on the DA$\Phi$NE 
high-level machine, reading and aligning the non-homogenous data 
from the distributed VME memories and writing them to the DA$\Phi$NE 
homogenous database;
\item the Global collector process (see Fig.~\ref{scheme}) 
on the KLOE slow control machine, reading data from the shared 
memories of low-level processes (via the SNMP inter-process protocol),
stored data from the high-level processes monitoring satellite systems, and
the physics quantities from the online event reconstruction.
\end{Itemize}
Some elaboration is also performed starting from low-level data, 
producing background estimators from spurious hits in the detector,
beam lifetimes from the fit of the history of the machine currents, etc.
 
Finally, the common global database is produced, in which all machine 
parameters are correlated with the detector quantities. Since there
is a wide range of time variations of monitored parameters and of
logging time intervals from the various processes (producing or
retrieving data), all the available quantities in the final common
database are synchronized and stored either in a 'fast file' (at
15 seconds intervals) or in a 'slow file' (at 1 minute intervals).

The global database is then accessed by
both the DA$\Phi$NE and KLOE controls Web servers to display a rich wealth
of information: 
\begin{Itemize}
\item as already described in the previous sections, the online
status; 
\item the long-term history of the machine and detector conditions; 
\item a number of statistics of the most relevant quantities.
\end{Itemize}
A number of pages displaying the status and the history of the various 
parameters and their correlations 
are made available and extensively used in both control rooms: 
mostly oriented to the luminosity/machine background optimization in the 
DA$\Phi$NE case, and to keep under control the detector performance 
and the data quality for KLOE. 


In order to help the two teams of physicists in the continous improvement
and optimization of the machine/detector operation, the presentation of
the monitored quantities and of their correlations is fundamental. 
This has been realized with two complementary tools taking
advantage of the common database:
a general-purpose Web interface, running on the DA$\Phi$NE Web server, 
mainly oriented to the display of correlations between different 
machine and detector
quantities, and a histogramming application based on the ROOT libraries 
(from the CERN program library), running on the KLOE online
machines~\cite{presenter}, mainly oriented to the presentation of the time 
charts of any machine or detector parameter. 
Both the presenting tools access the common database described above, and
can display the last few hours status or build the history on a many days
base.

An asynchronous builder process continously runs on the DA$\Phi$NE supervisor 
machine, reading the common global database: on the basis of the stored
parameters a number of statistics of the machine and detector performance 
are elaborated, such as delivered luminosity, beam lifetimes, data-taking
efficiency, etc.
The last-hours, daily or longer term
statistics of the most relevant indicators are then made available
through the Web interface.

As described above, the supervisor machine also runs a watchdog process
checking the most relevant parameters and alarm conditions. A dedicated
server takes care of signalling anomalous conditions through the e-mail
and GSM short message service, and in addition the daily statistics
can be broadcast to the authorized personnel.
This can be done in push/pull mode: the operators
of the two teams can get the desired information on-demand, or wait
for the regular updates.

For safety reasons the KLOE slow control also implements an independent
watchdog process, signalling the detector experts
the main alarm conditions via e-mail and GSM short messages. 

\section{Conclusions}
The DA$\Phi$NE and KLOE control systems manage and monitor the machine and
detector elements through a set of low-level programs. The high-levels
of the two systems are strongly connected and integrated: the information
coming at different times from the detector and machine controls, from
the external control systems, and from the online event reconstruction
on the KLOE farm are collected, synchronized, stored, pre-analyzed and
displayed from a number of tools, based on a common general database
accessed mainly by the two supervisor Web sites and by dedicated programs. 
Duplication is very limited, since the same low-level parameters are
elaborated and shown with different approaches, oriented either to machine 
optimization, or to detector stability and data quality control. 

\section{Acknowledgments}
We are grateful to M.~Masciarelli for the work on the DA$\Phi$NE
side, to A.~Balla and G.~Corradi for the work on the
KLOE side of the control system.

\end{document}